# Uncertainty-Aware Multi-Parametric Magnetic Resonance Image Information Fusion for 3D Object Segmentation

*Cheng Li[1,4], Yousuf Babiker M. Osman[1,2], Weijian Huang[1,2,3], Zhenzhen Xue[1,4], Hua Han[1,2], Hairong Zheng[1], Shanshan Wang[1,3,4]\**

[1]Paul C. Lauterbur Research Center for Biomedical Imaging, Shenzhen Institutes of Advanced Technology, Chinese Academy of Sciences, Shenzhen 518055, China.
[2]University of Chinese Academy of Sciences, Beijing 100049, China.
[3]Peng Cheng Laboratory, Shenzhen 518066, China.
[4]Guangdong Provincial Key Laboratory of Artificial Intelligence in Medical Image Analysis and Application, Guangzhou 510080, China.

**ABSTRACT**

Multi-parametric magnetic resonance (MR) imaging is an indispensable tool in the clinic. Consequently, automatic volume-of-interest segmentation based on multi-parametric MR imaging is crucial for computer-aided disease diagnosis, treatment planning, and prognosis monitoring. Despite the extensive studies conducted in deep learning-based medical image analysis, further investigations are still required to effectively exploit the information provided by different imaging parameters. How to fuse the information is a key question in this field. Here, we propose an uncertainty-aware multi-parametric MR image feature fusion method to fully exploit the information for enhanced 3D image segmentation. Uncertainties in the independent predictions of individual modalities are utilized to guide the fusion of multi-modal image features. Extensive experiments on two datasets, one for brain tissue segmentation and the other for abdominal multi-organ segmentation, have been conducted, and our proposed method achieves better segmentation performance when compared to existing models.

## 1. INTRODUCTION

Multi-parametric magnetic resonance (MR) imaging can obtain rich information from the imaged regions, and thus, it has been widely employed in clinical practice to help visualize and quantify different tissues and organs [1], [2]. Volume-of-interest (VOI) segmentation is necessary for image quantification. Manual VOI delineation is still the preferred approach in the clinic, which is labor-exhaustive and time-consuming. Errors could happen because of fatigue. Automated algorithms are highly desired to help radiologists achieve fast and accurate image inspection and quantification.

Deep learning has achieved unprecedented performance in multi-parametric MR image segmentation in recent years [3], [4]. There is one key factor in this process that is not fully explained: the effective fusion of information or features extracted from different modalities. Existing studies address this issue by direct pixel-wise feature summation or feature concatenation at different network layers. For example, Nie et al. added the deep features in their network for brain tissue segmentation [5]. Li et al. introduced a supervised image fusion method specifically for the task of breast tumor segmentation using dynamic contrast-enhanced MR imaging and T2-weighted MR imaging [6]. Dolz et al. proposed a densely connected convolutional neural network (CNN) for multi-modal MR image segmentation by densely concatenating the features from different modalities and different layers [3]. All these works have achieved impressive performances in their respective tasks. Nevertheless, it is still not clear how and why the fusion of the image information can help achieve enhanced segmentation performance. As a result, the generalizability of these existing methods is difficult to guarantee.

In this study, we propose an uncertainty-aware multi-parametric MR image information fusion method for 3D object segmentation. Instead of simple image feature summation or concatenation, we design a specific module that employs the uncertainties in individual modality predictions as a measure of the information entropy. We suggest that those image voxels in one modality that can give certain predictions should be highlighted and those give uncertain predictions might contain interfering information, the influence of which should be weakened. With this criterion, our proposed uncertainty-aware image feature fusion network can fuse the information of different modalities in a task-oriented manner, and enhanced segmentation performances in different tasks are expected.

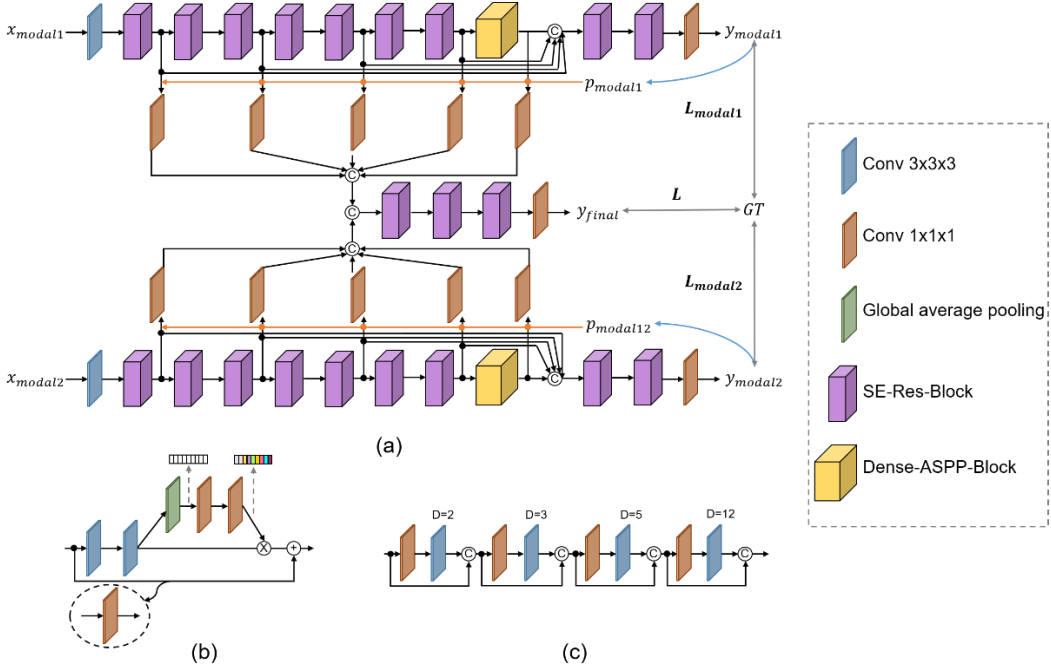

**Figure 1.** The proposed uncertainty-aware multi-parametric MR image information fusion network. (a) The overall network architecture. (b) SE-Res-Block in (a). (c) Dense-ASPP-Block in (a).

## 2. METHODOLOGY

To effectively extract and fuse the information provided by each imaging modality, an uncertainty-aware multi-parametric MR image information fusion network is proposed (Fig. 1). Here, the situation of two input modalities is plotted because the datasets experimented with in this study provide data of two MR imaging sequences. The network can be easily extended to situations with more than two input modalities.

Overall, two network streams are included, which are responsible for the information extraction from the input data. Each stream is self-contained, meaning that it can work independently to accomplish the segmentation task and generate the respective segmentation results ($y_{modal1}$ and $y_{modal2}$ in Fig. 1a). Both streams are made up of two major building blocks, SE-Res-Block (Fig. 1b) and Dense-ASPP-Block (Fig. 1c). SE-Res-Block combines the residual connections with the squeeze-and-excitation attention mechanism [7], [8]. In order to achieve accurate segmentation results on the tortuous boundaries of brain tissues as well as on small organs, no pooling operation is utilized in the network. Instead, Dense-ASPP-Block, which contains a series of dilated convolutions with different dilation rates, is introduced to enlarge the receptive field. In this way, the model can handle both large and small targets very well. Multi-level features are combined by direct concatenation to make modality-specific predictions ($y_{modal1}$ and $y_{modal2}$).

To fuse the extracted image features from the two modalities, an uncertainty-aware fusion module is proposed. Supposing $y_{modal1} \in [0, 1]^{M \times N \times D \times C}$ is the probability map generated by stream 1, where M, N, and D refer to the three dimensions of the 3D image data and C is the number of classes, the uncertainties in the prediction are calculated as: $U_{modal1} = (\prod_i (y_{modal1,i})) / (1/N)^N$, where $i$ represents the class ($i = 1, ..., N$). We suggest that the higher the uncertainty, the more interfering information the voxel has. To avoid the negative effects of the interfering information, actions need to be taken to weaken the influence of these voxels on the final segmentation. In implementation, we choose to simply multiply the extracted image features by $(1 - U_{modal1})$ to highlight the trustworthy voxels and weaken the ambiguous voxels. The same procedures are followed by stream 2, and the extracted features are multiplied by $(1 - U_{modal2})$. Then, the obtained uncertainty-aware multi-level image features are fused by concatenation after passing through the respective adaptation layers. The final prediction $y_{final}$ is made by utilizing the fused features.

In total, there are three outputs ($y_{modal1}$, $y_{modal2}$, and $y_{final}$). The entire network is trained in a two-stage manner. In the first stage, when epoch number is smaller than 30 (a hyper-parameter determined empirically), the loss function is calculated as: $L_{stage1} = 0.5 \times L_{ce}(y_{modal1}, GT) + 0.5 \times L_{ce}(y_{modal2}, GT)$, where $L_{ce}$ refers to the cross-entropy loss and GT stands for ground-truth segmentation maps. In the second stage, when epoch number is no smaller than 30, the loss function becomes: $L_{stage2} = 0.5 \times L_{ce}(y_{modal1}, GT) + 0.5 \times L_{ce}(y_{modal2}, GT) + L_{ce}(y_{final}, GT)$. This scheduled training can reduce the large errors caused by initial fluctuated predictions of stream 1 and stream 2.

## 3. RESULTS

Experiments on two datasets from two challenges, iSeg-2019 [2] and CHAOS [1], have been conducted to evaluate the effectiveness of the proposed method. Ethical approval was not required. There are 10 3D MR images (two sequences: T1-weighted (T1W) and T2-weighted (T2W) imaging) with provided labels for brain tissue segmentation (cerebrospinal fluid (CSF), gray matter (GM), and white matter (WM)) in iSeg-2019. Following the practice of one existing method, HyperDenseNet [3], we randomly divided the 10 images into a training set of 6 images, a validation set of 1 images, and a test set of 3 images. CHAOS provides 20 labeled 3D images (two imaging parameters: in-phased and opposed-phased MR imaging) for abdominal multi-organ segmentation (liver, right kidney, left kidney, and spleen). Similarly, these 20 images were randomly divided into 12 images for training, 2 images for validation, and 6 images for testing. Two comparison methods were implemented, a classical 3D U-Net model [9] and HyperDenseNet [3]. The experimental settings were the same for all the methods on both datasets. Image patches of $32 \times 32 \times 32$ were extracted with an extraction step of $14 \times 14 \times 14$. We only keep those meaningful image patches (with target) during training. Cross-entropy loss was calculated. All experiments were conducted using the PyTorch framework on a Tesla V100 GPU (32GB).

Table 1 lists the segmentation results on the iSeg-2019 dataset. It can be observed that our baseline model (Ours* in Table 1) can already generate better segmentation results than 3D U-Net and HyperDenseNet. The uncertainty-aware feature fusion module can further enhance the segmentation accuracy by increasing the Dice scores of segmenting all three brain tissues. On the other hand, segmentation with only the data from one modality, especially T2W, leads to decreased performance.

**Table 1.** Segmentation results (Dice: %) on the iSeg-2019 dataset (* means segmentation with our designed model but without the uncertainty-aware feature fusion module.)

| Method | CSF | GM | WM |
|---|---|---|---|
| 3D U-Net | 93.1±0.3 | 87.7±0.2 | 85.1±0.3 |
| HyperDenseNet | 93.0±0.0 | 87.6±0.1 | 85.1±0.0 |
| Ours* (T1W) | 93.8±0.1 | 89.2±0.1 | 86.4±0.0 |
| Ours* (T2W) | 88.5±0.0 | 83.3±0.2 | 80.7±0.0 |
| Ours* | 93.9±0.2 | 90.0±0.1 | 87.8±0.1 |
| Ours (T1W) | 93.8±0.1 | 89.3±0.1 | 86.4±0.1 |
| Ours (T2W) | 89.0±0.0 | 83.8±0.1 | 81.0±0.1 |
| **Ours** | **94.2±0.1** | **90.5±0.0** | **88.2±0.0** |

Segmentation results on the CHAOS dataset are given in Tables 2 (Dice scores for each class) and 3 (Dice scores averaged over the four classes). The results show that our proposed method achieves much better segmentation results than HyperDenseNet. One possible reason is that we include the Dense-ASPP-Block, which is able to increase the receptive field of the network. Regarding the uncertainty-aware feature fusion module, better overall segmentation performance is obtained when the module is enabled. The module is especially helpful for the segmentation of the right kidney and the spleen.

**Table 2.** Segmentation results (Dice: %) on the CHAOS dataset (# means a training image patch resample strategy is introduced to balance the samples over the four classes, which can solve the class-imbalance issue partially. All our methods utilize this strategy. * means segmentation with our model but without the uncertainty-aware fusion module.)

| Method | Liver | Right kidney | Left Kidney | Spleen |
|---|---|---|---|---|
| HyperDenseNet | 75.2±3.0 | 35.0±0.1 | 37.7±0.7 | 35.7±3.3 |
| HyperDenseNet# | 77.8±2.4 | 33.7±1.5 | 41.8±2.0 | 39.9±1.2 |
| Ours* (In) | 80.1±1.1 | 46.2±0.1 | 43.8±1.9 | 52.4±0.6 |
| Ours* (Opposed) | 82.1±1.6 | 37.4±2.5 | 43.1±0.1 | 65.3±1.7 |
| Ours* | **87.4±1.1** | 45.2±1.7 | **46.9±1.1** | 64.4±0.2 |
| Ours (In) | 79.1±1.0 | 46.9±1.0 | 40.0±6.2 | 53.4±1.3 |
| Ours (Opposed) | 78.9±3.7 | 38.2±1.6 | 43.8±2.1 | 66.3±5.2 |
| **Ours** | 86.9±0.4 | **47.9±1.6** | 45.7±1.2 | **67.5±2.8** |

**Table 3.** Average segmentation results (Dice: %) of different methods on the CHAOS dataset

| Method | Hyper DenseNet | Hyper DenseNet# | Ours* | **Ours** |
|---|---|---|---|---|
| Avg | 45.91 | 48.32 | 60.98 | **62.02** |

Example segmentation results of different methods on the two datasets are plotted in Fig. 2 and Fig. 3. More accurate segmentation results can be observed in the segmentation maps of our proposed methods, especially on the segmentation of abdominal organs.

## 4. CONCLUSION

In this study, we propose an uncertainty-aware information fusion method for the segmentation of multi-parametric MR images. Uncertainties in modality-specific predictions were utilized to guide the feature fusion of images acquired with different imaging parameters, making our network more task-oriented and more generalizable. Extensive experiments on two public datasets validated the effectiveness of the proposed method. Our method can be very helpful in clinical applications where multi-parametric MR imaging is widely adopted.

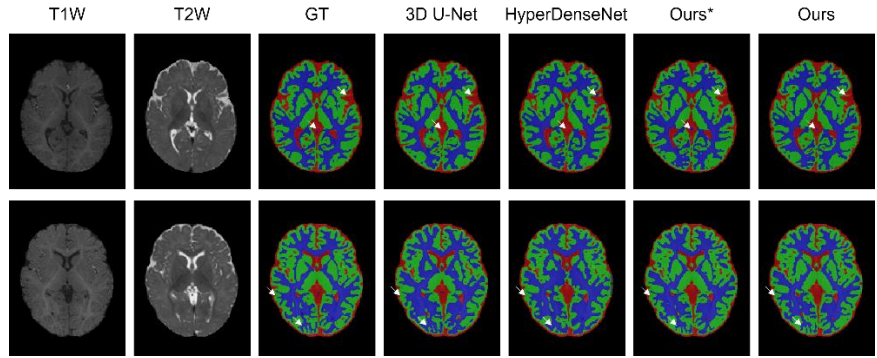

**Figure 2.** Example segmentation results of different methods on the iSeg-2019 dataset (GT shows the ground-truth segmentation map). Red, green, and blue color regions represent the segmentation results of CSF, GM, and WM. White arrows indicate the regions where our methods segment better than existing methods.

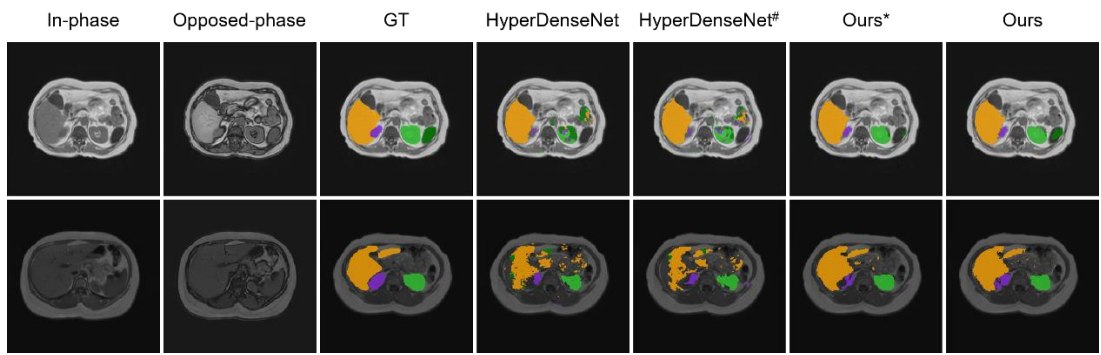

**Figure 3.** Example segmentation results of different methods on the CHAOS dataset. Yellow, purple, light green, and dark green regions indicate the segmentation results of the liver, the right kidney, the left kidney, and the spleen, respectively.

## 5. ACKNOWLEDGMENTS

This research was partly supported by Scientific and Technical Innovation 2030-"New Generation Artificial Intelligence" Project (2020AAA0104100, 2020AAA0104105), the National Natural Science Foundation of China (61871371), Guangdong Provincial Key Laboratory of Artificial Intelligence in Medical Image Analysis and Application (2022B1212010011), the Basic Research Program of Shenzhen (JCYJ20180507182400762), Shenzhen Science and Technology Program (JCYJ20220531100213029, RCYX20210706092104034), and Youth Innovation Promotion Association Program of Chinese Academy of Sciences (2019351).